\documentstyle[twoside,fleqn,espcrc2,epsfig]{article}

\newcommand{\ste}{\tilde{t}_1}
\newcommand{\mse}{m_{\tilde{t}_1}}
\newcommand{\steb}{\bar{\tilde{t}}_1}

\newcommand{\st}{\tilde{t}}

\newcommand{\sq}{\tilde{q}}
\newcommand{\gt}{\tilde{g}}

\newcommand{\ct}{\tilde{\chi}}
\newcommand{\et}{\tilde{l}}

\newcommand{\mg}{m_{\tilde{g}}}
\newcommand{\mq}{m_{\tilde{q}}}

\def\simgt{\rlap{\lower 3.5 pt \hbox{$\mathchar \sim$}} \raise 1pt \hbox {$>$}}
\def\simlt{\rlap{\lower 3.5 pt \hbox{$\mathchar \sim$}} \raise 1pt \hbox {$<$}}

\newcommand{\plb}[3]{Phys.\ Lett.\ {\bf B#1} (19#2) #3}

\newcommand{\npb}[3]{Nucl.\ Phys.\ {\bf B#1} (19#2) #3}
\newcommand{\prd}[3]{Phys.\ Rev.\ {\bf D#1} (19#2) #3}

\newcommand{\prl}[3]{Phys.\ Rev.\ Lett.\ {\bf #1} (19#2) #3}

\newcommand{\AmS}{{\protect\the\textfont2
  A\kern-.1667em\lower.5ex\hbox{M}\kern-.125emS}}

\title{Supersymmetric particle production at hadron colliders}

\author{Michael Kr\"amer\address{CLRC Rutherford Appleton Laboratory,
        Chilton, Didcot, OX11 0QX, England\\[0.5em] {\small Talk
        presented at the International Euroconference on Quantum
        Chromodynamics (QCD 98), Montpellier, France, 2-8 Jul 1998 and
        at the SUSY 98 Conference, Oxford, UK, 11-17 Jul 1998, RAL
        preprint RAL-TR-1998-065.}}}

\begin{document}

\begin{abstract}
  The theoretical status of MSSM particle production at the hadron
  colliders Tevatron and LHC is reviewed, including next-to-leading
  order supersymmetric QCD corrections.  The higher-order corrections
  significantly reduce the theoretical uncertainty and lead to a rise
  of the lower bounds on supersymmetric particle masses, as
  demonstrated for the case of top-squark and gaugino pair production
  at the Tevatron.
\end{abstract}

\maketitle

\section{Introduction}
The search for supersymmetric particles ranks among the most important
experimental endeavours at existing and future hadron colliders. At
the upgraded $p\bar{p}$ collider Tevatron, chargino and neutralino
searches, as well as squark and gluino searches, will cover a wide
range of the \mbox{MSSM} parameter space \cite{CCEFM-97}. Squarks and
gluinos up to masses $\simlt$~2.5~TeV can be discovered at the $pp$
collider \mbox{LHC} \cite{LHC}, so that the entire canonical parameter
space of strongly interacting particles in low-energy SUSY will be
explored eventually.

The cross sections for the production of SUSY particles in hadron
collisions have been calculated at the Born level already quite some
time ago \cite{LO}.\footnote{The MSSM Higgs sector will not be
discussed here, see e.g.\ Refs.\cite{Higgs} for recent reviews.}  Only
recently have the theoretical predictions been improved by
calculations of the next-to-leading order SUSY-QCD corrections for the
production of
\\[1mm]
\begin{tabular}{ll}
 $-$ squarks, gluinos & 
 $p\bar{p}/pp \to \sq\bar{\sq},\gt\gt,\sq\gt$ \cite{BHSZ-95}\\[1mm]

 $-$ top-squark pairs & 
 $p\bar{p}/pp \to \st\bar{\st}$ \cite{BKPSZ-98}\\[1mm]

 $-$ gaugino pairs & 
 $p\bar{p}/pp \to \ct\ct$ \cite{BKKPSZ-98,LHC-talks,TP-98}\\[1mm]

 $-$ slepton pairs & 
 $p\bar{p}/pp \to \et\et$ \cite{BHR-98}\\[2mm]

\end{tabular}
The higher-order corrections in general increase the production cross
section compared to the predictions at the Born level and thereby
improve experimental mass bounds and exclusion limits. Moreover, by
reducing the dependence of the cross section on spurious parameters,
{\it i.e.} the renormalization and factorization scales, the cross
sections in NLO are under much better theoretical control than the
leading-order estimates.

In the simplest realization of supersymmetric grand unified theories,
the lightest supersymmetric particles are in general the non-coloured
gauginos, with masses in the range 50 to 200~GeV for the lightest of
these states.  Within the strongly interacting SUSY sector, the
top-squark (stop) eigenstate $\ste$ is expected to be the lightest
particle \cite{ER-83}. Stop and gaugino pair production are therefore
among the most promising reactions for supersymmetric particle
searches at the Tevatron and will be focussed upon in the following.

\section{Top-squark production}

\begin{figure}[htb]
 \vspace*{-2cm}
 \hspace*{1.25cm}
 \epsfysize=17.cm
 \epsfxsize=12.cm
 \centerline{\epsffile{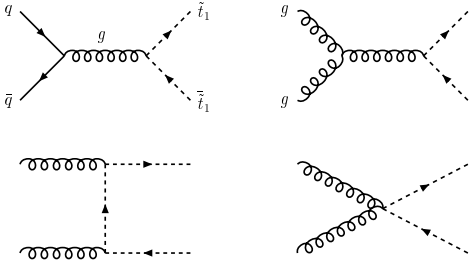}} 
\vspace{-11.5cm}
\caption[ ]{\it  Generic leading-order Feynman diagrams for the production of stop
 pairs in quark-antiquark annihilation and gluon fusion.}
\label{fig:stop-diag}
\vspace*{-5mm}
\end{figure}

At hadron colliders, the lowest order QCD processes for the production
of $\ste$ pairs are quark--antiquark annihilation and gluon--gluon
fusion
\begin{equation}
q \bar{q} \to \ste \steb \quad \mbox{and} \quad 
g g       \to \ste \steb
\label{eq:stop}
\end{equation}
as shown in Fig.~{\ref{fig:stop-diag}}.
The hadronic $p\bar{p}/pp$ cross sections are obtained by folding the
partonic cross sections with the $q\bar{q}$ and $gg$ luminosities.  At
the Tevatron the dominant mechanism for large stop masses
$\mse\;\simgt\; 100$~GeV is the valence $q\bar{q}$ annihilation. The
fraction of $q\bar{q}$ initiated events rises from 50 to 80\%, if the
$\ste$ mass is increased from 100 to 200~GeV. At the LHC the
gluon-fusion mechanism plays a more prominent role. For a $\ste$ mass
below 200~GeV, more than 90\% of the events are generated by $gg$
fusion.

\begin{figure}[htb]
 \vspace*{-6mm}
 \epsfysize=7.0cm
 \epsfxsize=7.8cm
 \centerline{\epsffile{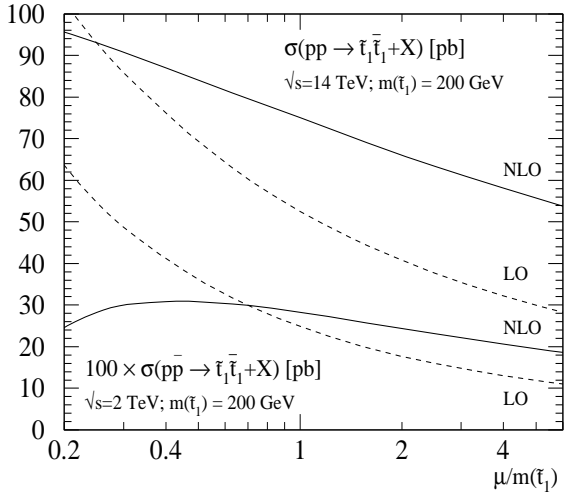}} 
\vspace{-1cm}
\caption[ ]{\it  The renormalization/factorization-scale dependence of the total
  cross section for $\tilde{t}_1$-pair production at the Tevatron and
  the LHC.}
\label{fig:stop-scale}
\vspace*{-5mm}
\end{figure}

The lowest order cross section depends strongly on the renormalization
and factorization scales. As a result, the theoretical predictions are
uncertain within factors of two. Including the SUSY-QCD corrections
\cite{BKPSZ-98,TP-98}, this scale dependence is reduced significantly,
as shown in Fig.~\ref{fig:stop-scale}.  At the same time the cross
section is considerably enhanced at the central scale ($\mu = \mse$).

\begin{figure}[htb]
  \vspace*{-6mm}
  \epsfysize=7.0cm 
  \epsfxsize=7.8cm 
  \centerline{\epsffile{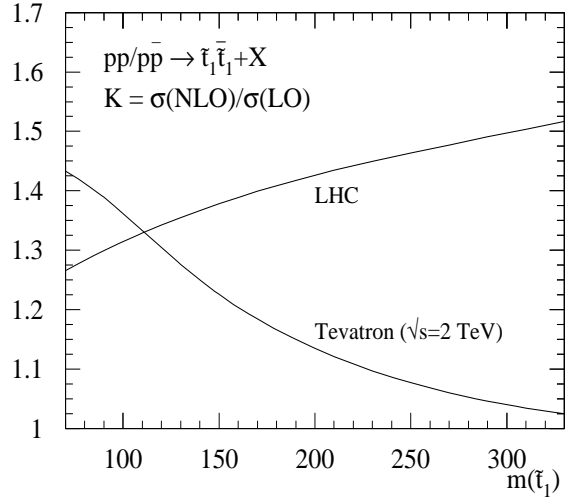}}
  \vspace{-1cm}
\caption[ ]{\it $K$-factors for $\tilde{t}_1$-pair production
    at the Tevatron and the LHC as a function of the stop mass.  The
    renormalization and factorization scale is fixed at the central
    value $\mu = \mse$.}
\label{fig:stop-K}
\vspace*{-7.5mm}
\end{figure}

The magnitude of the SUSY-QCD corrections is illustrated by the $K$
factors in Fig.~\ref{fig:stop-K} for $\ste$-pair production at the
Tevatron and the LHC. The $K$ factor is defined as
$K\!=\!\sigma_{NLO}/\sigma_{LO}$, with all quantities calculated
consistently in lowest and in next-to-leading order. In the mass range
considered, the SUSY-QCD corrections are positive and reach a level of
30 to 50\% if the $gg$ initial state dominates. If, in contrast, the
$q\bar{q}$ initial state dominates, the corrections are small.  The
relatively large mass dependence of the $K$ factor for $\ste\steb$
production at the Tevatron can therefore be attributed to the fact
that the $gg$ initial state is important for small $\mse$, whereas the
$q\bar{q}$ initial state dominates for large $\mse$.

The cross section for $\ste$-pair production depends essentially only
on the stop mass $\mse$. The dependence of the cross section on the
other SUSY parameters, {\it i.e.}~the gluino mass, the masses of the
other squarks and the mixing angle, which enter at next-to-leading
order, is very weak. The squark and gluino contributions in virtual
loops are essentially decoupled for canonical SUSY masses and the
general behaviour of the higher-order corrections is determined by
ordinary QCD gluon radiation.  Bounds on the $\ste\steb$ production
cross section can therefore easily be translated into lower bounds on
the lightest stop mass without reference to other supersymmetric
parameters. On the other hand, if stop particles were to be
discovered, the cross section can be exploited directly to determine
the stop mass $\mse$. The impact of the SUSY-QCD corrections on the
experimental mass bounds for $\ste$ will be discussed in
Sec.~\ref{sec:mass-bounds}.

\section{Gaugino production}

In many supersymmetric models, the light neutralinos and charginos are
significantly lighter than squarks or gluinos, and may therefore be
the only supersymmetric particles directly accessible at the
Tevatron. Because of the low SM background, the production of
$\tilde{\chi}^{0}_{2} \tilde{\chi}^{\pm}_{1} $ followed by the decays
$\tilde{\chi}^{0}_{2} \to \tilde{\chi}^{0}_{1} l^{+} l^{-} $ and
$\tilde{\chi}^{\pm}_{1} \to \tilde{\chi}^{0}_{1} l^{\pm} \nu$ is one
of the gold-plated trilepton SUSY signatures at colliders.

\begin{figure}[htb]
 \vspace*{-2.9cm}
 \hspace*{1.25cm}
 \epsfysize=17.cm
 \epsfxsize=12.cm
 \centerline{\epsffile{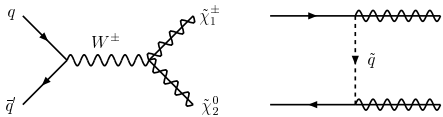}} 
\vspace{-14cm}
\caption[ ]{\it  Generic leading-order Feynman diagrams for the
 production of $\tilde{\chi}^{0}_{2} \tilde{\chi}^{\pm}_{1}$ pairs in
 quark-antiquark annihilation and $t$-channel squark exchange.}
\label{fig:gaugino-diag}
\vspace*{-7mm}
\end{figure}

Neutralinos and charginos can be produced at hadron colliders in
quark-antiquark annihilation via $s$-channel gauge boson
$(\gamma,Z,W)$ production and $(t,u)$-channel squark exchange, e.g.\
\begin{equation}
u \bar{d} \to \tilde{\chi}_{2}^{0}\tilde{\chi}_{1}^{+}
\quad \mbox{and} \quad 
d \bar{u} \to \tilde{\chi}_{2}^{0}\tilde{\chi}_{1}^{-}
\label{eq:gaug}
\end{equation}
as shown in Fig.~{\ref{fig:gaugino-diag}}.
The production cross sections are not simple functions of the chargino
and neutralino masses but depend strongly on the mixing (i.e.\ on the
gaugino and Higgsino compositions) and the squark masses. In the
following, for illustration the supersymmetric parameters have been
fixed within the minimal supergravity model \cite{DM-95} taking the
GUT scale parameters at $m_{1/2}=150$~GeV, $m_0=100$~GeV,
$A_0=300$~GeV, $\tan\beta=4$, and $\mu >0$. The sparticle masses in
this scenario are given by $m_{\tilde{\chi}_{2}^{0}}=103$~GeV,
$m_{\tilde{\chi}_{1}^{\pm}}=100$~GeV, $\mg=402$~GeV, $\mq=352$~GeV and
$\ste=197$~GeV.

\vspace*{1mm}

\begin{figure}[htb]
 \vspace*{-5mm}
 \epsfysize=7.0cm
 \epsfxsize=7.8cm
 \centerline{\epsffile{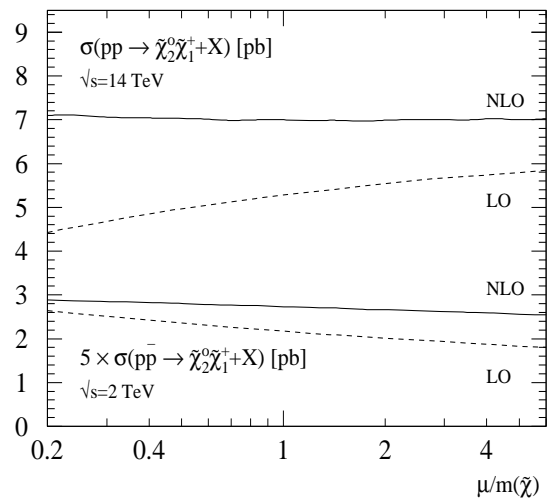}} 
\vspace{-1.1cm}
\caption[ ]{\it 
  The renormalization/factorization-scale dependence of the total
  cross section for $\tilde{\chi}^{0}_{2} \tilde{\chi}^{+}_{1}$
  production at the Tevatron and the LHC. $m_{\tilde{\chi}}$ denotes
  the average mass of the produced particles.}
\label{fig:gaugino-scale}
\vspace*{-6mm}
\end{figure}

The hadronic gaugino-pair production cross section depends on the
factorization scale via the parton densities. This scale dependence is
reduced when the SUSY-QCD corrections \cite{BKKPSZ-98,LHC-talks,TP-98}
are included. The renormalization scale dependence introduced by the
${\cal O}(\alpha_s)$ corrections is weak and the overall theoretical
uncertainty due to scale variation is small at the NLO level, see
Fig.~\ref{fig:gaugino-scale}.

\begin{figure}[htb]
% \vspace*{-7.5mm}
 \epsfysize=7.0cm
 \epsfxsize=7.8cm
 \centerline{\epsffile{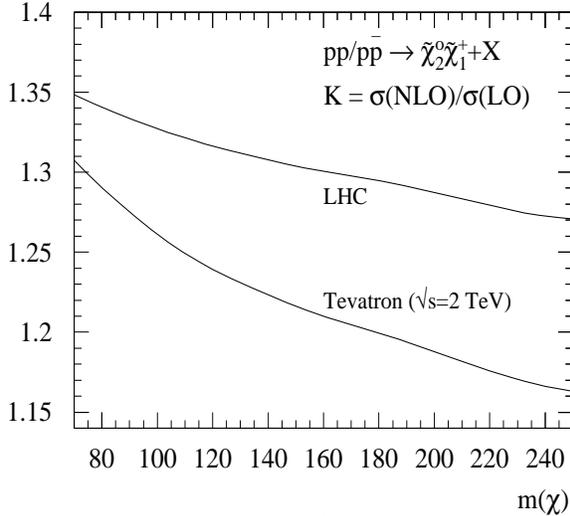}} 
\vspace{-1.2cm}
\caption[ ]{\it 
  $K$-factors for $\tilde{\chi}^{0}_{2} \tilde{\chi}^{+}_{1}$
  production at the Tevatron and the LHC as a function of the gaugino
  mass $m_{\tilde{\chi}}$ (the masses of $\tilde{\chi}^{0}_{2}$ and
  $\tilde{\chi}^{\pm}_{1}$ differ only marginally in the SUGRA
  inspired scenario adopted here). The cross sections are varied with
  $m_{1/2}$.  The renormalization and factorization scale is fixed at
  the average final state particle mass $\mu = m_{\tilde{\chi}}$.}
\label{fig:gaugino-K}
\vspace*{-7.5mm}
\end{figure}

The SUSY-QCD corrections significantly enhance the cross section for
the production of gaugino pairs at the Tevatron and the LHC, as
illustrated by the $K$ factors in Fig.~\ref{fig:gaugino-K}. For the
$\tilde{\chi}^{0}_{2} \tilde{\chi}^{\pm}_{1}$ cross at the Tevatron,
the higher-order corrections reach a level of $\sim$ 30\% for small
$m_{1/2}$ (and, accordingly, small gaugino masses) and decrease with
increasing $m_{1/2}$. At the LHC the corresponding $K$ factors are
larger and range between 1.25 and 1.35. A detailed analysis of all
relevant gaugino pair production cross sections at the Tevatron and
the LHC will be given in Ref.~\cite{BKKPSZ-98}.

\section{Experimental searches and mass bounds}\label{sec:mass-bounds}
\begin{figure}[htb]
% \vspace*{-7.5mm}
 \epsfysize=7.0cm
 \epsfxsize=7.8cm
 \centerline{\epsffile{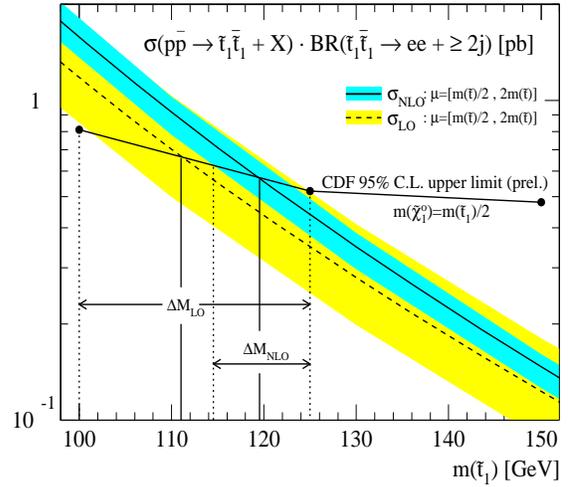}} 
\vspace{-1cm}
\caption[ ]{\it 
  The 95\% confidence level cross section times branching ratio upper
  limit for $\ste\steb$ production at the Tevatron \cite{CDF-stop}
  compared to the LO and NLO theoretical prediction as a function of
  the stop mass $\mse$. The theoretical curves are multiplied by a
  branching ratio of 1/8. The LO and NLO bands show the
  renormalization and factorization scale dependence of the cross
  section prediction.}
\label{fig:stop-mass-bounds}
\vspace*{-10mm}
\end{figure}

SUSY particle searches have been performed by the \mbox{CDF} and
\mbox{D0} collaborations at the Fermilab Tevatron.  No signal has been
found in the analysed data samples, resulting in upper cross section
limits \cite{exp-searches}. Comparing the experimental limits with the
theoretical cross section prediction one can exclude parts of the SUSY
parameter space or place lower bounds on the sparticle masses. The
impact of the higher-order QCD corrections on the interpretation of
the experimental cross section limits is exemplified in
Fig.~\ref{fig:stop-mass-bounds} for stop searches in like-sign
dielectron plus multijet events \cite{CDF-stop} which can arise in
$R$-parity violating neutralino decays $p \bar{p} \to \ste \steb \to c
\tilde{\chi}^{0}_{1} \bar{c} \tilde{\chi}^{0}_{1} \to e^{\pm}e^{\pm}+ 
\ge 2j$. The SUSY-QCD corrections increase the lower mass bound
by $\sim 10$~GeV and, in particular, they reduce the theoretical
uncertainty by a factor $\sim 4$. Qualitatively similar conclusions
hold for $\ste$ searches in $R$-parity conserving decay modes
\cite{CDF-D0-stop}.

\section{Summary}
\begin{figure}[h]
 \vspace*{-6mm}
 \epsfysize=6.1cm
 \epsfxsize=7.8cm
 \centerline{\epsffile{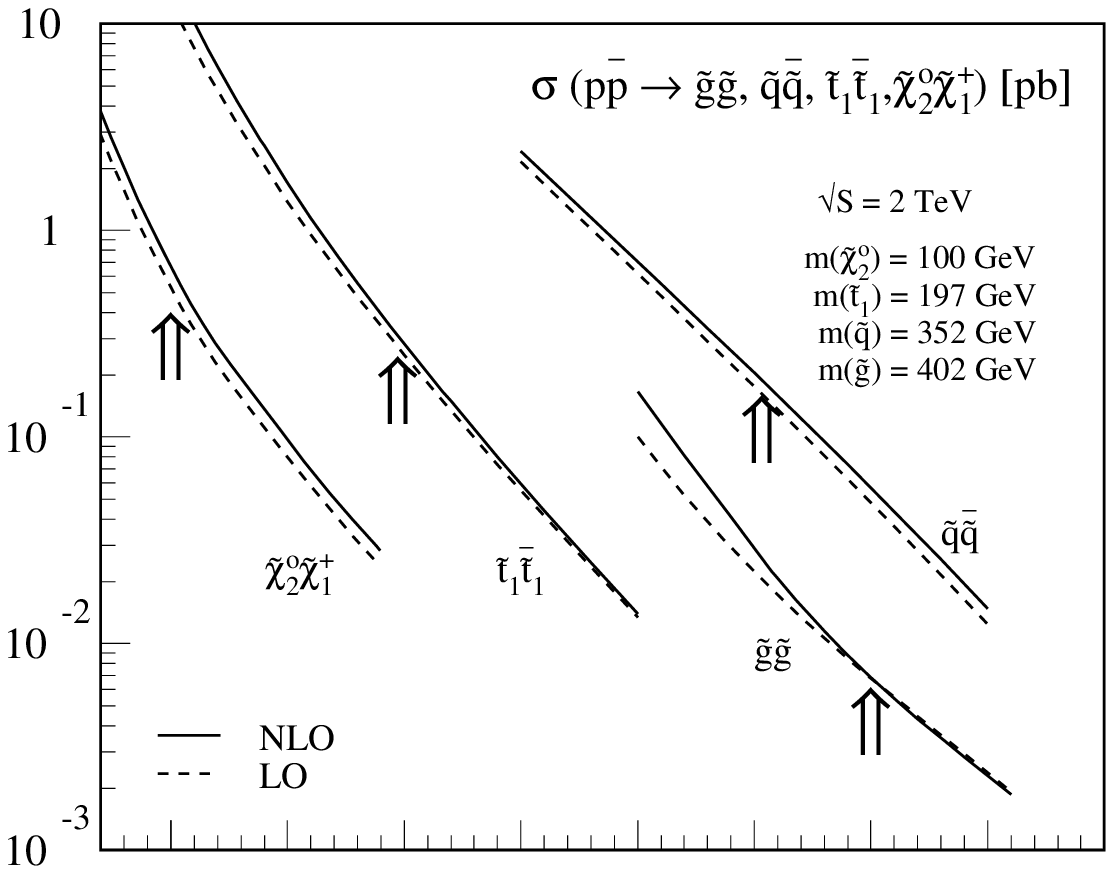}} 
 \vspace*{-0.5cm}
 \epsfysize=6.8cm
 \epsfxsize=7.8cm
 \centerline{\epsffile{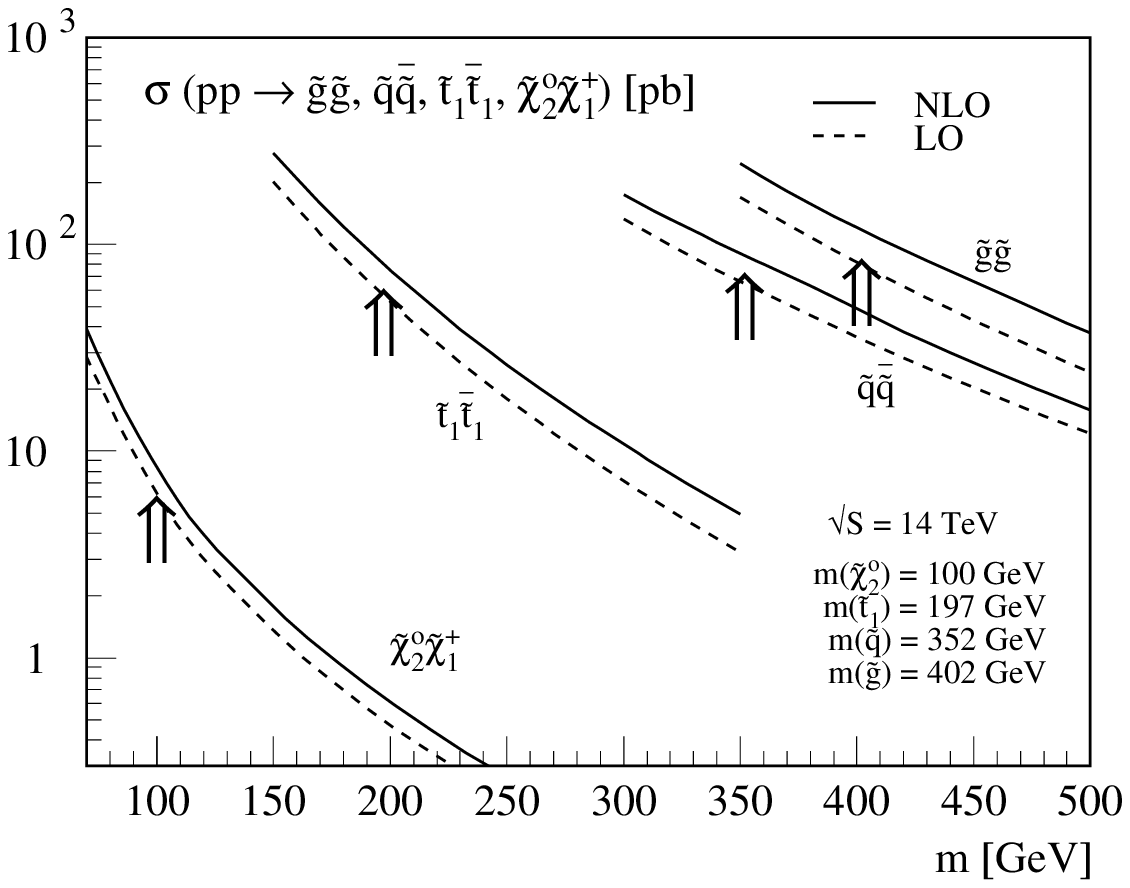}} 
\vspace{-1.2cm}
\caption[ ]{\it 
  The total cross section for pair production of squarks, gluinos,
  stops and gauginos as a function of the mass of the produced
  particles. The arrows denote the mass spectrum of the SUGRA inspired
  scenario described in the text. The cross sections are shown for the
  upgraded Tevatron and the LHC.}
\label{fig:sum}
\vspace*{-6mm}
\end{figure}

The next-to-leading order SUSY-QCD corrections for the production of
top-squark and gaugino pairs at the hadron colliders Tevatron and LHC
have been discussed. By reducing the scale dependence of the cross
section considerably, the quality of the theoretical predictions is
substantially improved compared with the lowest-order calculations. A
collection of relevant MSSM particle production cross sections at the
upgraded Tevatron and the LHC is shown in Fig.~\ref{fig:sum},
including next-to-leading order SUSY-QCD corrections.

The improved cross section predictions play a crucial role in the
experimental analyses. They either serve to extract the exclusion
regions in the SUSY parameter space from the data, or, in the case of
discovery, they can be exploited to determine the masses of the
sparticles.

\pagebreak

\noindent
{\small {\bf Acknowledgements}\\
 It is a pleasure to thank W.~Beenakker, M.~Klasen, T.~Plehn, M.~Spira
 and P.M.~Zerwas for their collaboration.}


\begin{thebibliography}{9}
\bibitem{BHR-98} The QCD corrections to slepton pair production at
  hadron colliders can be obtained from the well known results for SM
  Drell-Yan production, see H.~Baer, B.W.~Harris, and M.~Hall~Reno,
  \prd{57}{98}{5871}. The full supersymmetric QCD corrections will be
  presented in Ref.~\cite{BKKPSZ-98}.
\end{thebibliography}
\end{document}